\documentclass[
    ,final            
  ]
  {aipproc}

\layoutstyle{6x9}

\usepackage{natbib}
\usepackage{graphics}
\usepackage{epsfig}
\usepackage{amssymb}
\usepackage{graphicx}

\begin{document}

\title{Young Cool Stars Divided on the Issue of Rotation}


\author{S{\o}ren Meibom}{
  address={Harvard-Smithsonian Center for Astrophysics, 60 Garden Street, Cambridge, MA, 02138}
}



\begin{abstract}

We present the results of a combination of new stellar rotation periods
and extensive information about membership in the young open clusters
M\,35 and M\,34. The observations show that late-type members of both
M\,35 and M\,34 divide into two distinct groups, each with a different
dependence of rotation on mass (color). We discuss these new results in
the context of existing rotation data for cool stars in older clusters,
with a focus on the dependence of rotation on mass and age. We mention
briefly tests of rotation as an ``astronomical clock'' (gyrochronology),
and our plans to use the Kepler space mission to push observations of
stellar rotation periods beyond the age of the Hyades and the Sun.

\end{abstract}

\maketitle


\section{Introduction}

Understanding the evolution of stellar rotation is an integral part of
understanding stellar evolution. Clusters of stars allow us to probe
their rotational evolution as a function of their most fundamental
properties - mass and age. For cool stars (here: $0.5 < M_{\odot} < 1.25$)
in young open clusters, rotational periods can be measured to very
high precision. Combined with cluster membership information, such
observations now reveal clear relations between stellar rotation,
age, and mass.

Early observations of cool stars in open clusters younger than the
Hyades discovered that they rotate with periods ranging over two orders
of magnitude - from near breakup to periods similar to the Sun
\citep[see e.g.][and references therein]{va82,sh87,ssm+93}. This early
information was primarily from spectroscopic observations giving
projected rotational velocities ($v\sin i$), affected by the $\sin i$
ambiguity and the need to determine the stellar radius to derive
the angular rotation velocity. Despite these ambiguities, some structure
in the distribution of $v\sin i$ with stellar color was noted, as was
an emerging time- and mass-dependence for the presence of ultra fast
rotators (UFR). UFR were primarily found among the K and M dwarfs in
$\alpha$ Persei and the Pleiades, while lacking altogether in the older
Hyades cluster. These observations prompted ideas about the processes
in cool stars responsible for their rotational evolution, such as 
the suggestion of epochs of decoupling and recoupling of the stellar
core and envelope \citep{shs+84,ssm+93}.

Recently, from a study of rotational period data for several young open
clusters, Barnes (2003) \cite[][hereinafter B03]{barnes03a} proposed
that stars fall along two ``rotational sequences'' in the color vs.
period plane. From an analysis of these sequences and their dependencies
on stellar age, B03 proposed a framework for connecting internal and external
magneto-hydrodynamic processes to explain the evolution in the observed
period distributions. This approach combines the ideas of initial
decoupling of the stellar core and envelope, with re-connection of
the two zones through a global dynamo-field at a later and mass-dependent
time. B03 also proposed that the evolution of the rotational sequences
in the color-period diagram, could be used to measure the age of a stellar
population, much like the sequences in the color-magnitude diagram.
A similar idea had been proposed by \citet{kawaler89}. \citet{barnes07}
further developed this idea of ``gyrochronology''.

Interpretation aside, it is desirable and increasingly possible to
eliminate the ambiguities of $v\sin i$ data by measuring rotational
periods from light modulation by star-spots. It has also become more
feasible to carry out comprehensive surveys for kinematic membership
in star clusters. Combining the two reveals new details about dependencies
of rotation on the most fundamental stellar properties - mass and age.

\begin{figure}
\includegraphics[height=.35\textheight]{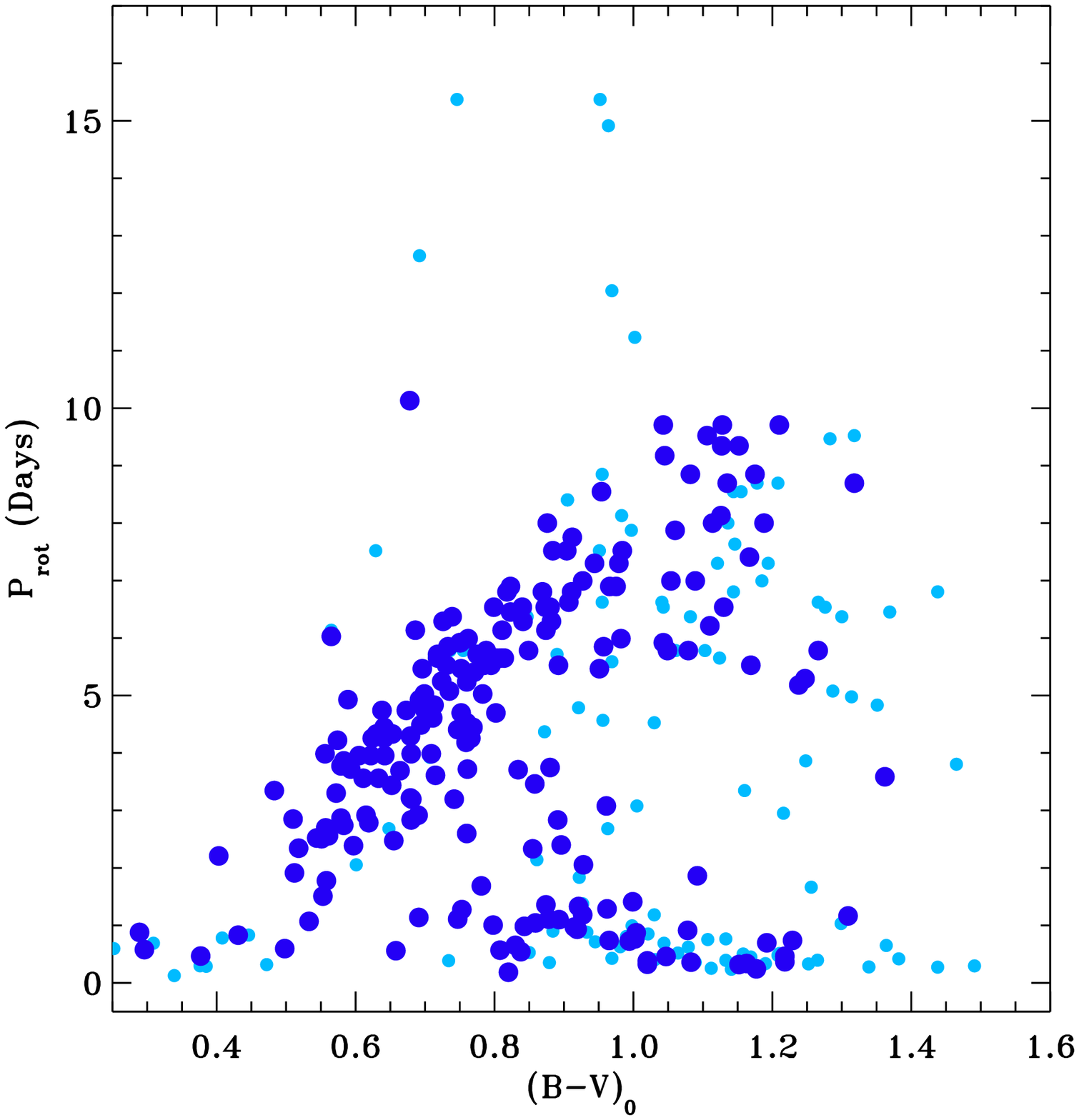}
\includegraphics[height=.35\textheight]{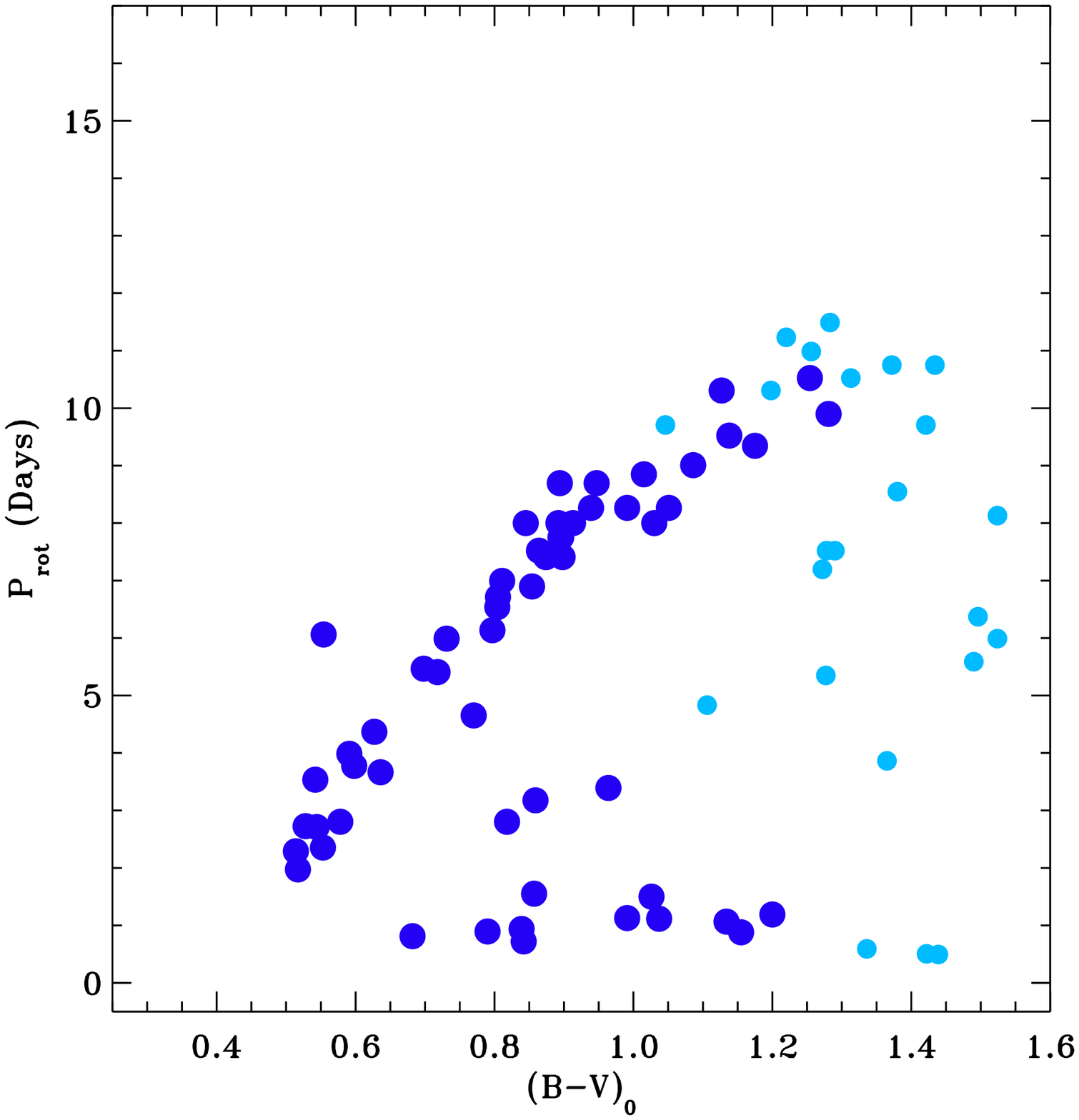}
\caption{The distribution of stellar rotation periods with (B-V)
color index for 310 members of M\,35 ({\it left}; \citep{mms08}) and
79 members of M\,34 ({\it right}). Dark blue plotting symbols are used
for radial-velocity members and light blue for photometric members.}
\label{m3534pbv}
\end{figure}


\section{New observations of two young clusters}

We carried out photometric monitoring campaigns over 5 consecutive months
for rotational periods, and nearly decade-long radial-velocity surveys for
cluster membership and binarity, on the 150\,Myr and 200\,Myr open clusters
M\,35 and M\,34. For full descriptions of the observations, data-reduction,
and data-analysis, see \citet{mm05,mms06,mms08}, and \citet{bmm08}.

{\it Time-Series Photometric Observations}:
We surveyed, over a timespan of 143 days, a region of $40 \times 40$ arc
minutes centered on each cluster. Images were acquired at a frequency of
once a night both before and after a central block of 16 full nights with
observations at a frequency of once per hour. The data were obtained in
the Johnson V band with the WIYN 0.9m telescope on Kitt Peak. Instrumental
magnitudes were determined from Point Spread Function photometry. Light
curves were produced for more than 14,000 stars with $12 < V < 19.5$.
Rotational periods were determined for 441 and 120 stars in the fields
of M\,35 and M\,34, respectively (see Figure~\ref{m3534pbv}).

{\it The spectroscopic surveys}:
M\,35 and M\,34 have been included in the WIYN Open Cluster Study
(WOCS; \citet{mathieu00}) since 1997 and 2001. As part of WOCS, 1-3
radial-velocity measurements per year were obtained on both clusters
within the 1-degree field of the WIYN 3.5m telescope with the multi-object
fiber positioner (Hydra) feeding a bench mounted echelle spectrograph.
Observations were done at central wavelengths of 5130\AA\ or 6385\AA\
with a wavelength range of $\sim$200\AA\ . From this spectral region with
many narrow absorption lines, radial velocities were determined with a
precision of $< 0.4~$km/s \citep{gmh+08,mbd+01}. Of the stars
with measured rotational periods in M\,35 and M\,34, 203 and 56, respectively,
are radial-velocity members of the clusters (dark blue symbols in
Figure~\ref{m3534pbv}). Including photometric members (light blue
symbols in Figure~\ref{m3534pbv}), the total number of stars with
measured rotational periods in M\,35 and M\,34, are 310 and 79, respectively.


\section{The color-period diagram}

Figure~\ref{m3534pbv} shows the rotational periods in M\,35 and M\,34
plotted against $B-V$ color. The coeval stars fall along two well-defined
sequences representing two different rotational states. One sequence
displays a clear correlation between period and color, and forms a
diagonal band of stars whose periods are increasing with increasing
color index (decreasing mass). The second sequence consists of UFR
and shows little mass dependence. Finally, a small subset of stars is
distributed between the two sequences. The distribution of stars in
the color-period diagrams suggests that the rotational evolution is
slow where we see the sequences, and fast in the gap between them,
while other areas of the color-period plane are either unlikely or
``forbidden''.

\begin{figure}
\includegraphics[height=.35\textheight]{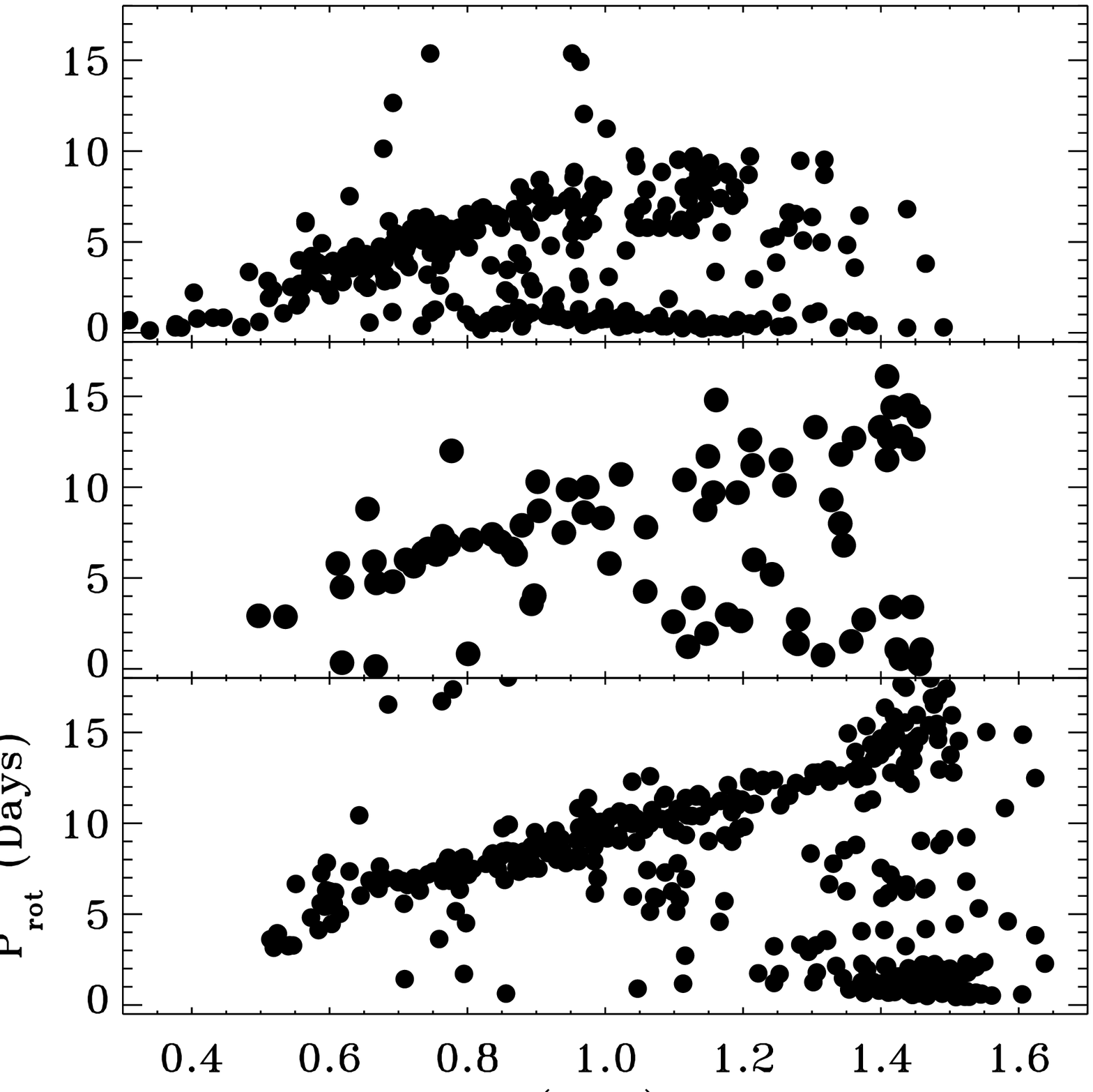}
\includegraphics[height=.35\textheight]{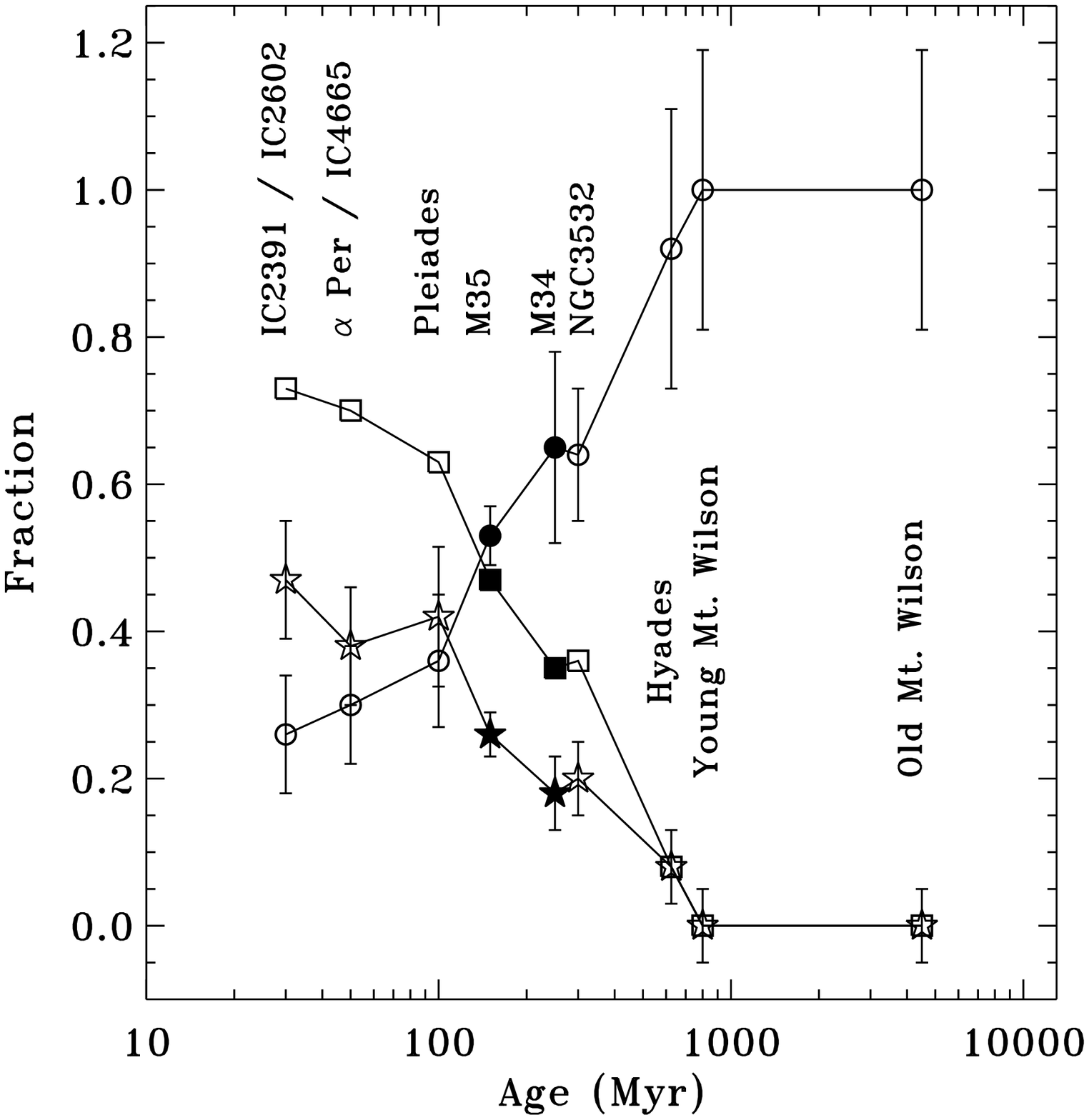}
\vspace{36pt}
\caption{{\it Left}: The color-period diagrams for clusters M\,35
(150\,Myr; top), NGC\,3532 (300\,Myr; center), and M37 (550\,Myr; bottom).
{\it Right}: Figure 3 from B03 with M\,35 and M\,34 added. The figure shows
the fractions of stars with $0.5 \le (B-V)_0 \le 1.5$ on the I sequence
(circles), the C sequence (stars), and C sequence and gap stars (squares),
for clusters of different ages.}
\label{3pbv}
\end{figure}

While much more apparent in M\,35 and M\,34, and in the study by
\citet{hgp+08} of the 550\,Myr cluster M37 (see below), the sequences
seen in Figure~\ref{m3534pbv} are the same as identified by B03.
B03 named the diagonal sequence the {\it interface} (I) sequence,
and the sequence of UFR the {\it convective} (C) sequence. Barnes
argues that stars on the C sequence have decoupled radiative cores
and convective envelopes, and suggests that the evolution of their
surface rotation is governed primarily by the moment of inertia
of the envelope and by inefficient loss of angular momentum linked
to small-scale convective magnetic fields. For stars on the I sequence,
B03 suggests that large-scale (sun-like) magnetic fields, produced
by an interface dynamo, provide more efficient Skumanich-style
angular momentum loss. B03 suggests that a late-type star evolves
from the C sequence and onto the I sequence when rotational shear
between the stellar core and envelope establish a large-scale dynamo
field. This happens sooner in higher mass stars as they have thinner
convective envelopes with smaller moments of inertia.

\subsubsection{The evolving color-period diagram: timescales, and
mass-dependence}

Color-period diagrams for clusters of different ages allow us to
study the rotation of cool stars as a function of mass and age.
In Figure~\ref{3pbv} (left) we show the color-period diagrams for
M\,35 (150\,Myr), NGC\,3532 (B03; 300\,Myr), and M37 (550\,Myr).
This diagram reveals both stellar spin-down on the I sequence and
a mass-dependence of the timescale for ``migration'' from the C
to the I sequence. At 150\,Myr (M\,35) G dwarfs have evolved onto
the I sequence, by 300\,Myr (NGC\,3532) the early K dwarfs have
followed, and by 550\,Myr (M\,37) most of the late K dwarfs are
on the I sequence as well. In Figure~\ref{3pbv} (right), we add
M\,35 and M\,34 to Fig. 3 in B03 which shows the fractions of cluster
stars on the I and C sequences as a function of their ages. M\,35
and M\,34 fit well the evolutionary trends of a decreasing fraction
of C sequence (and gap) stars and an increasing fraction of I sequence
stars with age. The almost linear trends suggest an exponential change
in time of the number of stars on the sequences. By counting stars
on both sequences and in the gap in the M\,35 color-period diagram,
and making the assumption that all stars start on the C sequence at
the ZAMS, we estimate the characteristic timescale for the rotational
evolution of stars off the C sequence and onto the I sequence. We
derive $\tau_c^G = 60$ Myr and $\tau_c^K = 140$ Myr for G and K dwarfs,
respectively. Such timescales may offer valuable constraints on the
evolution of stellar dynamos and internal and external angular momentum
transport.

\begin{figure}
\includegraphics[height=.35\textheight]{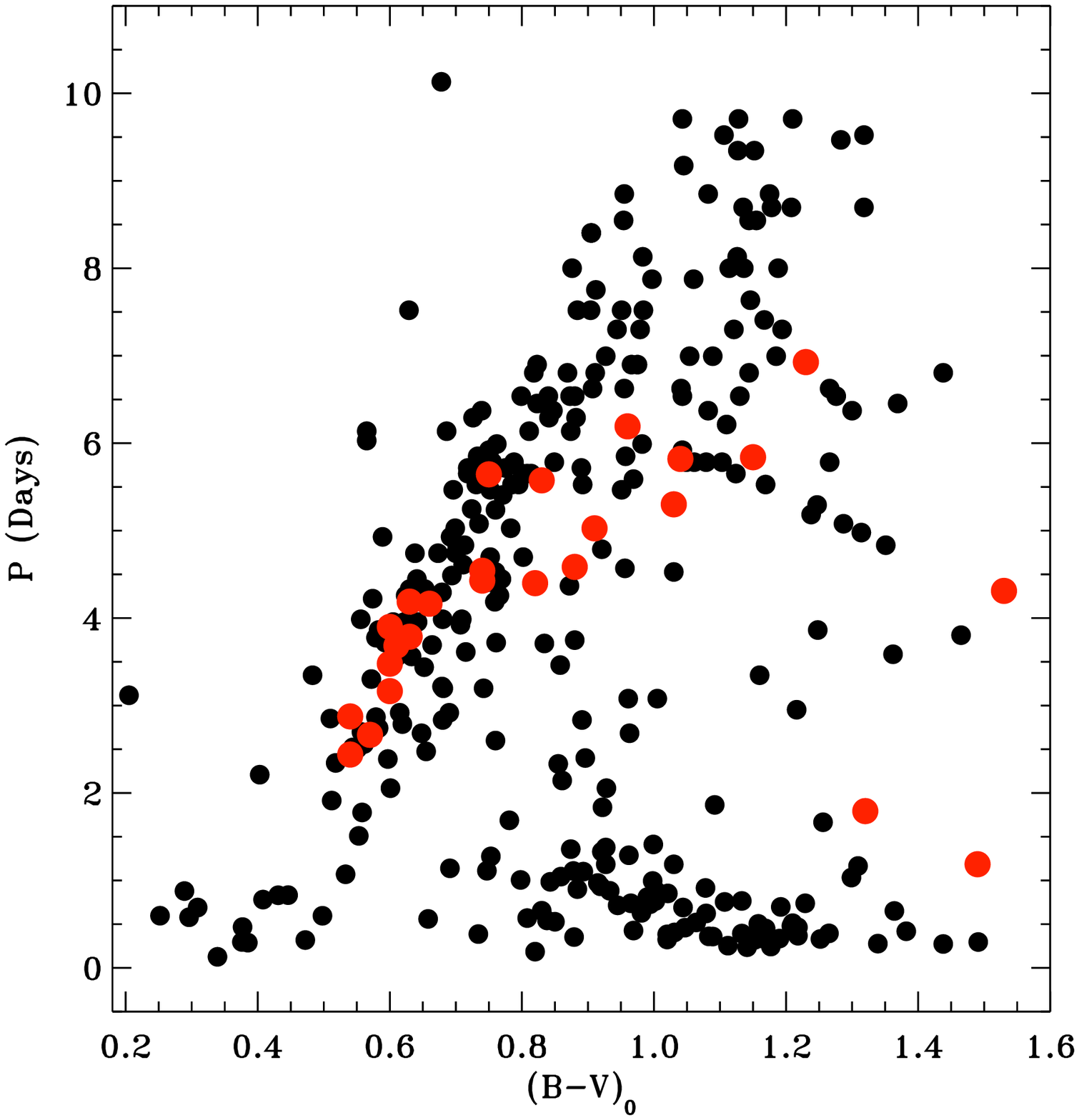}
\includegraphics[height=.35\textheight]{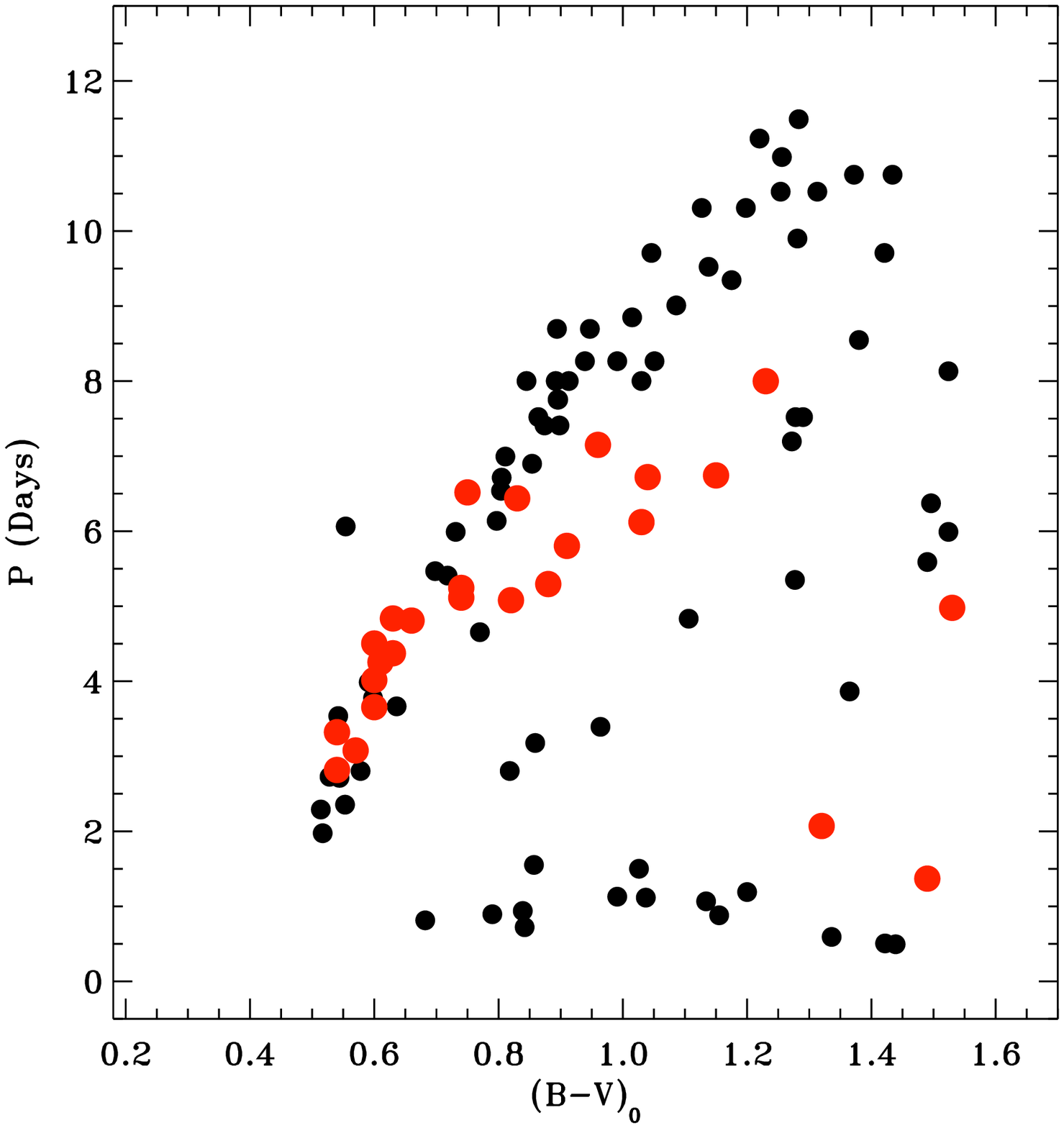}
\caption{The M\,35 ({\it left}) and M\,34 ({\it right}) color-period
diagrams with 25 Hyades stars overplotted (red symbols). In both plots,
the Hyades rotation periods were spun-up by factors of $\sqrt{625/150}$
and $\sqrt{625/200}$, respectively, in accordance with the Skumanich
$\sqrt{t}$ time-dependence on stellar rotation evolution, and an age
of 625\,Myr for the Hyades \citep{pbl+98}.}
\label{m3534pbv_hyades}
\end{figure}

\subsubsection{Testing the Skumanich relationship ($P_{rot} \propto
\sqrt{age}$)}

On the I sequence, angular momentum loss ($dJ/dt$) is thought to be
proportional to the angular rotation velocity ($\Omega$) to the third
power ($dJ/dt \propto \Omega^{3}$). This relationship has been used
in the most recent models of angular momentum evolution for late-type
stars \citep[e.g.][]{pkd90,cdp95,bs96,kpb+97,spt00}. Its origin is
in magnetized stellar wind theory \citep[][and references therein]
{kawaler88}. However, assumptions were made about the geometry of
the global magnetic field ($B$), and about the relation between $B$
and $\Omega$, in order to satisfy the Skumanich relationship ($P_{rot}
\propto \sqrt{age}$). We note that the Skumanich relationship was
derived based on $v\sin i$ data for solar-like stars (G dwarfs) in
open clusters and for the Sun. We ask here if the Skumanich relationship
is also valid for stars of lower mass?

Figure~\ref{m3534pbv_hyades} shows color-period diagrams in which the
M\,35 and M\,34 stars are plotted together with Hyades stars. The Hyades
stars have been spun-up in accordance with the Skumanich $\sqrt{t}$
time-dependence. Comparing stars on the I sequence, the $\sqrt{t}$
dependence appears to represent well the rotational evolution of the
G dwarfs between 150\,Myr/200\,Myr and 625\,Myr, whereas the spun-up
Hyades K-dwarfs have rotation periods systematically shorter than the
M\,35 and M\,34 K dwarfs. This suggests that the time-dependence on
spin-down of K dwarfs is slower than $\sqrt{t}$ in the age-interval
studied. A more detailed study of deviations from the Skumanich
relationship is underway.

\subsubsection{Testing gyrochronology}

\begin{figure}
\includegraphics[height=.35\textheight]{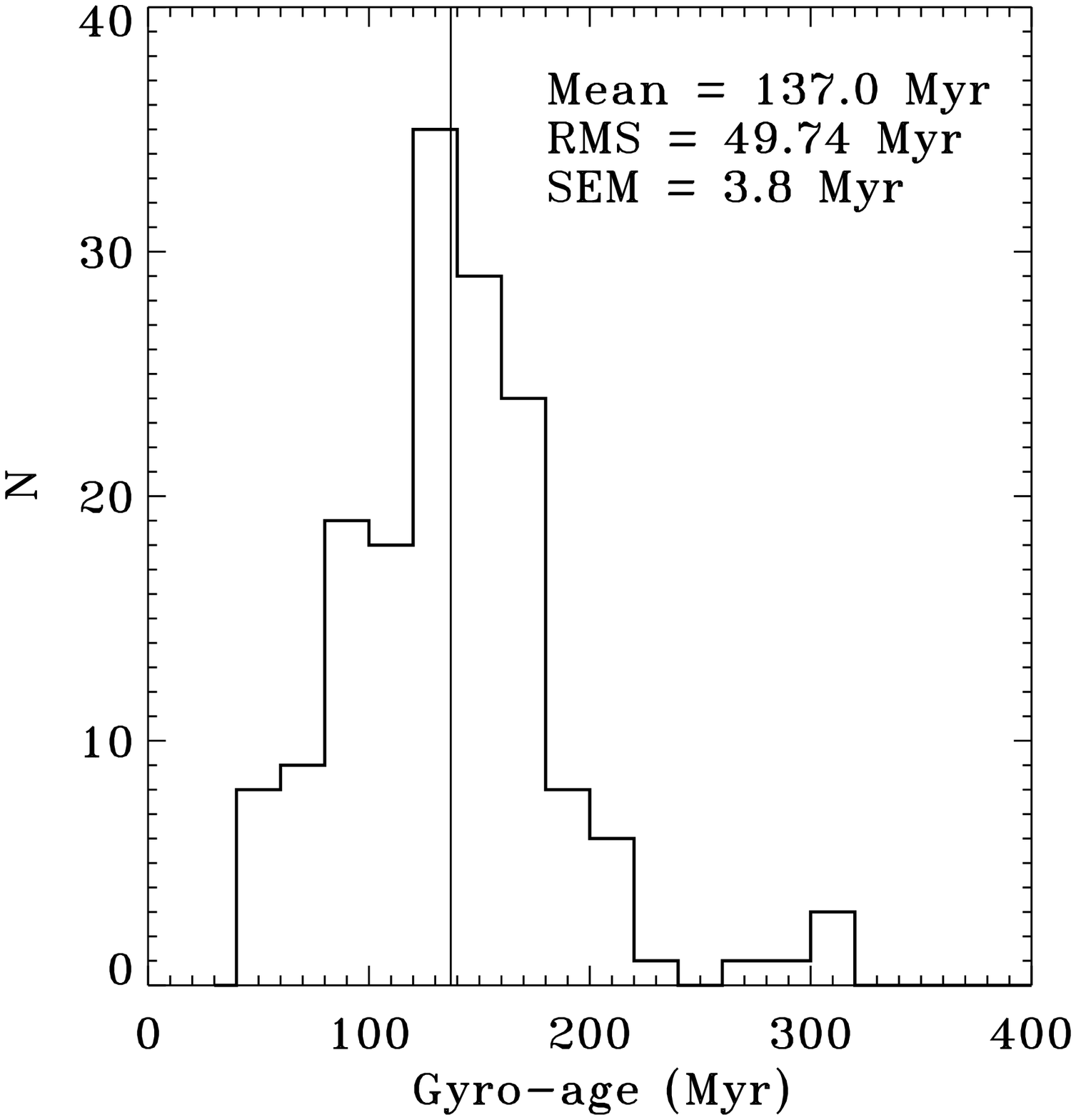}
\includegraphics[height=.35\textheight]{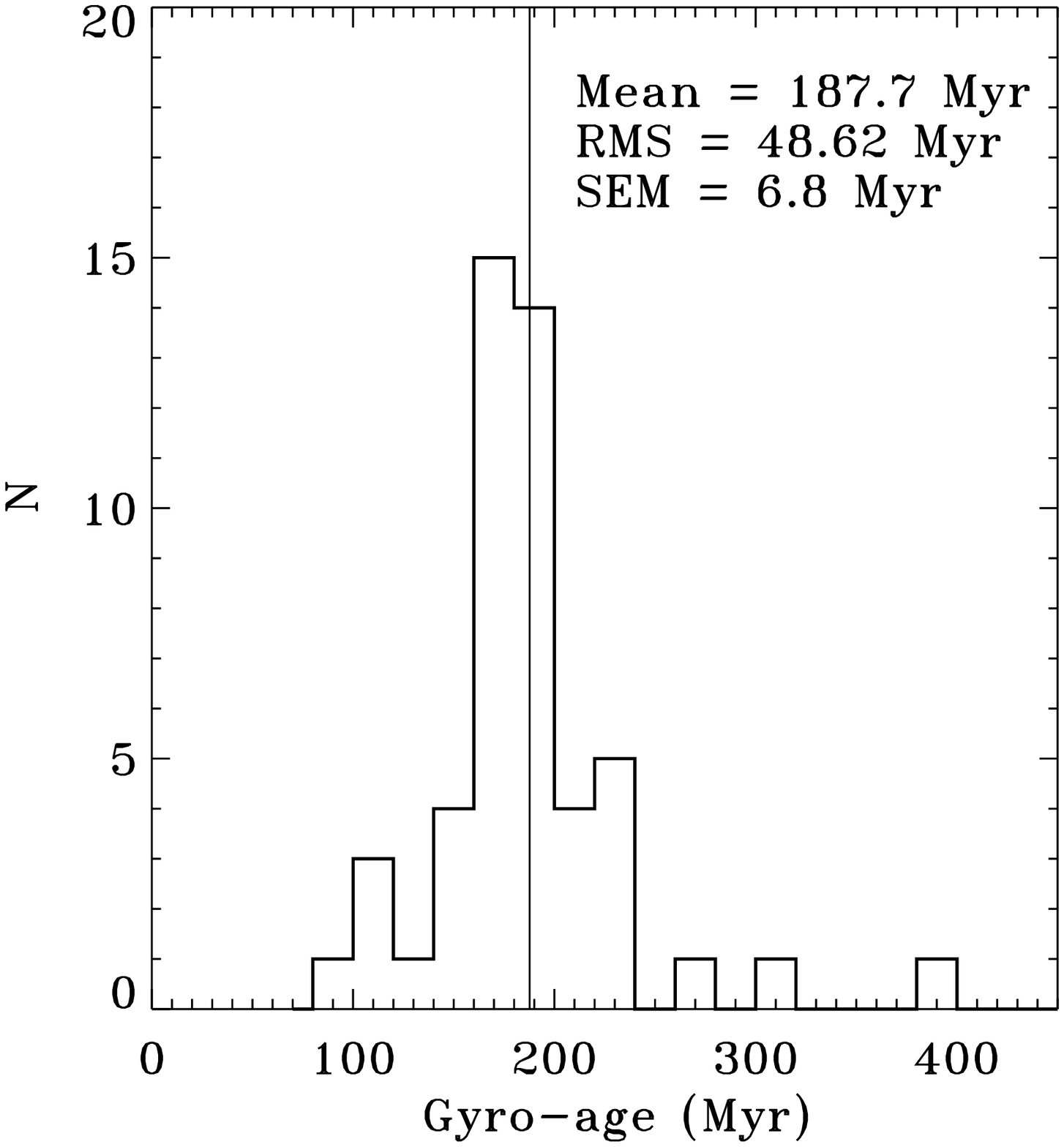}
\caption{The distributions of stellar ages for I sequence stars in M\,35
(left) and M\,34 (right) calculated using the B03 empirical age-rotation-color
relationship.
}
\label{gyroages}
\end{figure}

The idea of using the relations between stellar rotation, color (mass),
and age, to determine the latter from observations of the former two,
was proposed by \citet{kawaler89} and B03 (gyrochronology). Each proposed
an expression relating the age of a star to its rotation period and $B-V$
color. We show in Figure~\ref{gyroages}, the distributions of ages calculated
using the expression in B03 for I sequence stars in M\,35 and M\,34.
The mean gyro-ages for M\,35 and M\,34 are 137\,Myr and 188\,Myr. The
corresponding mean gyro-ages using the \citet{kawaler89} expression
are 161\,Myr and 217\,Myr. Both sets of gyro-ages fall within the range
of the isochrone ages determine for the two clusters (see \citet{mms08}
for a more detailed comparison and discussion).

\subsubsection{The Kepler mission - a unique opportunity}

The Kepler space mission - a search for transiting earth-like planets -
will provide time-series photometry of unprecedented duration, cadence,
and precision. The location of 4 open clusters within the Kepler field
of view, with ages from $\sim$0.5\,Gyr till $\sim$8\,Gyr, offers a unique
opportunity to measure rotational periods for cool stars beyond the age
of the Hyades and the Sun, and thereby extent the age-dimension of studies
of stellar angular momentum evolution. The author is leading an effort
to study the clusters with Kepler.

\begin{theacknowledgments}
We thank the organizers for a wonderful meeting.
\end{theacknowledgments}



\bibliographystyle{aipproc}   

\bibliography{refs}


\end{document}